\documentclass[aps,pra,twocolumn,american,showkeys,showpacs,floatfix,reprint]{revtex4-1}
\usepackage{lmodern}
\usepackage[T1]{fontenc}
\usepackage[koi8-r,latin9]{inputenc}
\usepackage{geometry}
\usepackage{units}
\usepackage{amsmath}
\usepackage{amssymb}
\usepackage{graphicx}
\usepackage{setspace}

\makeatletter
 
 \@ifundefined{textcolor}{}
 {%
   \definecolor{BLACK}{gray}{0}
   \definecolor{WHITE}{gray}{1}
   \definecolor{RED}{rgb}{1,0,0}
   \definecolor{GREEN}{rgb}{0,1,0}
   \definecolor{BLUE}{rgb}{0,0,1}
   \definecolor{CYAN}{cmyk}{1,0,0,0}
   \definecolor{MAGENTA}{cmyk}{0,1,0,0}
   \definecolor{YELLOW}{cmyk}{0,0,1,0}
 }

\makeatother

\usepackage{babel}
\begin{document}

\title{The Born Rule and Time-Reversal Symmetry \\
of Quantum Equations of Motion}

\author{Aleksey V. Ilyin}

\email{a.v.ilyin@mail.mipt.ru}

\affiliation{Moscow Institute of Physics and Technology}

\date{\today}
\begin{abstract}
It was repeatedly underlined in literature that quantum mechanics
cannot be considered a closed theory if the Born Rule is postulated
rather than derived from the first principles. In this work the Born
Rule is derived from the time-reversal symmetry of quantum equations
of motion. The derivation is based on a simple functional equation
that takes into account properties of probability, as well as the
linearity and time-reversal symmetry of quantum equations of motion.
The derivation presented in this work also allows to determine certain
limits to applicability of the Born Rule.
\end{abstract}



\maketitle

\section{Introduction\label{sec:Introduction}}

The Born Rule, being one of the basic principles of quantum physics,
establishes a link between a solution of wave equation and the results
of observations. The Born Rule was proposed in 1926 on the grounds
that it complies with the conservation of the number of particles
in scattering process \cite{Born}. Later, the Born Rule was supplemented
by the von~Neumann projection postulate \cite{vNeumann} to describe
the results of repeated measurements.

From the beginning, the Born Rule was perceived as an axiom independent
of quantum equations of motion. There were numerous attempts to derive
the Born Rule from the basic quantum principles. For example, a derivation
of the Born Rule was discussed in \cite{Deutsch} and later in \cite{Zurek-2003,Zurek-2005,Zurek-2011}
within the many-worlds interpretation of quantum mechanics. However,
this approach was criticized in~\cite{Critical-2008}. The Born Rule
was obtained in \cite{Aerts} within the concept of hidden measurements,
which, however, is not a widely accepted interpretation of quantum
mechanics. Recently, the Born Rule was derived in \cite{Lesovik}
from the unitarity of quantum evolution and an additional assumption
about probability conservation.

In this paper, the Born Rule is derived from the basic quantum principles,
namely, time-reversal symmetry and linearity of quantum equations
of motion. The derivation presented here is based on the following
premisses:

($\alpha$) The state of a particle is completely determined by its
wave function (WF), which is a solution of quantum equation of motion,
for example, the Schroedinger equation;

($\beta$) It follows from the above principle that all properties
of the particle, including the probability density~$P$, shall be
determined solely by the wave function $\psi$ of the particle, i.\,e.~$P=P(\psi)$.

Our purpose is to find function~$P(\psi)$.

\section{\label{sec:Derivation}Derivation of the Born Rule}

Probability density $P(\psi)$ is assumed to be a \emph{universal}
function that is applicable to any state of quantum particle described
by the~WF. Therefore, it will be sufficient to find $P(\psi)$ in
some simple special case, for example, in the case of a plane wave,
while the solution found in this special case, due to its universality,
is sure to be valid for an arbitrary~WF. By probability density of
finding a particle in a plane wave we mean the probability that the
particle is in a unit volume.

\subsection{\label{sub:Amplitudes}Amplitudes}

Consider a plane wave with complex amplitude~$S$ propagating along
$x$-axis: 
\begin{equation}
\psi(x,t)=S\exp i(kx-wt),\label{eq:Plane_wave}
\end{equation}
where $k$ and $w$ are proportional, respectively, to particle momentum
and energy $k=\tfrac{p}{\hbar}$, $w=\tfrac{E}{\hbar}$.

\begin{figure}[ht]
\noindent \begin{centering}
\includegraphics[width=8.5cm]{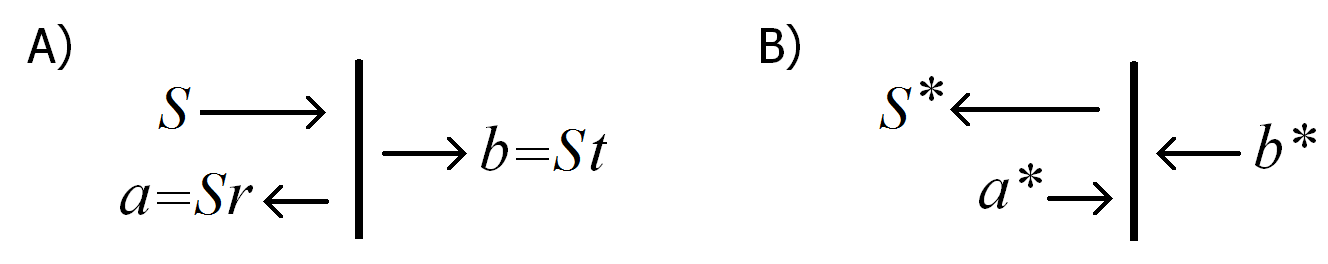} 
\par\end{centering}

\caption{\label{fig:Barrier}{\small Direct (A) and time-reversed (B) processes
of particle interaction with the barrier. Complex amplitudes $a$
of reflected wave and $b$ of transmitted wave become complex-conjugates
$a^{*}$ and $b^{*}$ under time reversal (see explanation in the
text.)}}
\end{figure}

Let a plane wave \eqref{eq:Plane_wave} be incident on a potential
barrier characterized by complex reflectance $r$ and transmittance~$t$
(labeling transmittance and time by the same symbol is traditional
and may cause no confusion.) Then the reflected wave and transmitted
wave (Fig.~\ref{fig:Barrier}A) will have complex amplitudes, respectively,
\begin{equation}
a=Sr\,\,\,\,\,\,\,\mathrm{and}\,\,\,\,\,\,\, b=St.\label{eq:a&b}
\end{equation}

It is well known \cite{Landau} that the Schroedinger equation is
symmetric with respect to time reversal if, along with the substitution
$t\rightarrow-t$, the WF is changed to its complex conjugate (see
Appendix for a simple explanation). 

Let the simultaneous action of time reversal and complex conjugation
be denoted as~$^{\sim}$, i.\,e. $\widetilde{\psi}(x,t)=\psi^{*}(x,-t)$.
This operation yields a time-reversed state of quantum particle. In
the general case, time-reversibility of the WF should be regarded
as a formal mathematical property of quantum equations of motion.
That property does not imply that any state of quantum particle can
actually be reversed in an experiment. 

Thus, reversing quantum state \eqref{eq:Plane_wave} yields 
\begin{equation}
\widetilde{\psi}(x,t)=S^{*}\exp i(-kx-wt).\label{eq:Plane_wave_reversed}
\end{equation}
Therefore, the reversed plane wave has a complex-conjugate amplitude~$S^{*}$.

Now let us reverse the process of interaction of a particle with the
barrier. Two waves with amplitudes $a$ and~$b$ that were departing
from the barrier now will go back to the barrier and their complex
amplitudes will turn into conjugate amplitudes $a^{*}$ and $b^{*}$
respectively (Fig.~\ref{fig:Barrier}B). Assuming the transmittance
and reflectance of the potential barrier being the same in either
direction (that is certainly true for a symmetrical barrier), one
obtains the amplitude of the wave reflected to the left as $ra^{*}$
and transmitted to the left as~$tb^{*}$. When time is reversed,
a wave of amplitude~$S^{*}$ should appear running away from the
barrier (Fig.~\ref{fig:Barrier}B.)

It follows from the linearity of quantum equations of motion that
the superposition principle holds for the~WF. This conclusion applies
to both direct and time-reversed processes. Therefore, the reversed
state $S^{*}$ is the result of superposition of two waves running
to the left from the barrier, namely, the reflected wave $ra^{*}$
and transmitted wave~$tb^{*}$: 
\begin{equation}
ra^{*}+tb^{*}=S^{*}.\label{eq:s}
\end{equation}
 Exponential factors $\exp i(-kx-wt)$ of all waves cancel in this
equation.

Multiplying \eqref{eq:s} by $S$ and taking into account \eqref{eq:a&b}
one obtains 
\begin{equation}
aa^{*}+bb^{*}=SS^{*},\label{eq:ss}
\end{equation}
which is an inevitable consequence of time-reversibility of quantum
equations and the superposition principle.

\subsection{\label{sub:Probabilities}Probabilities}

A particle falling on the barrier is \emph{indivisible}, so that it
can be found with probability $P(a)$ in the reflected beam or with
probability $P(b)$ in the transmitted beam (Fig.~\ref{fig:Barrier}A).
As these events are mutually exclusive (i.\,e. a particle cannot
be found both in the reflected and transmitted beams) the probabilities
should obey 
\begin{equation}
P(a)+P(b)=P(S).\label{eq:P(s)}
\end{equation}

Probability $P(\psi)$ must be a real number that, according to Provision
($\beta$), should be obtained from complex~WF. Real $x$ can be
obtained from complex $\psi$ using infinite number of ways. For example,
one may take a real part of complex number $x=Re(\psi)$, or complex
number argument $x=Arg(\psi)$, or complex number modulus $x=\left|\psi\right|=\sqrt{\psi\psi^{*}}$.
Also, various combinations of above expressions will produce real
numbers. 

In order to select a feasible expression for real number $P(\psi)$
from among the multitude of expressions, the following physical principle
should be applied: probability $P(\psi)$ should be independent of
arbitrarily chosen time origin because the properties of infinite
plane wave do not depend on time. 

Only expression $\psi\psi^{*}$ will have this property while other
expressions mentioned above will explicitly depend on time through
the WF argument~$kx-\omega t$. Therefore, with no loss of generality
one may assume that $P(\psi)$ is some function of argument~$x=\psi\psi^{*}$.
In other words, 
\begin{equation}
P(\psi)=F(\psi\psi^{*}).\label{eq:F(aa)}
\end{equation}
Such a designation is justified because it imposes no additional restrictions
on the unknown function~$F(x)$. Therefore, equation (\ref{eq:P(s)})
can be written as 
\begin{equation}
F(aa^{*})+F(bb^{*})=F(SS^{*}).\label{eq:P(ss)}
\end{equation}
According to Provision ($\beta$), the probability can be a function
only of the WF amplitude, which is provided by equation~\eqref{eq:P(ss)}.

\subsection{\label{sub:Functional-Equation}The Born Rule}

Due to \eqref{eq:ss} equation \eqref{eq:P(ss)} can be written as
\begin{equation}
F(aa^{*})+F(bb^{*})=F(aa^{*}+bb^{*}).\label{eq:Functional_Eq}
\end{equation}
That is a functional equation with respect to the unknown function~$F(x)$.
Equation \eqref{eq:Functional_Eq} can be presented in the usual form:
\begin{equation}
F(x)+F(y)=F(x+y),\label{eq:Final_eq.}
\end{equation}
where $x=aa^{*}$, $y=bb^{*}$.

It is important to note that here $x$ and $y$ are independent variables
because the height of the barrier (and its transparency) can be altered
independently of amplitude $S$ of the incident wave. Therefore, the
only solution to functional equation \eqref{eq:Final_eq.} is a linear
function 
\begin{equation}
F(x)=kx=kaa^{*},\label{eq:Linear}
\end{equation}
 which in view of \eqref{eq:F(aa)} yields 
\begin{equation}
P(a)=kaa^{*}.\label{eq:Born_Rule}
\end{equation}
Constant $k$ is actually a normalization factor that defines the
WF unit of measurement. In each quantum problem this normalization
factor is chosen based on convenience. For example, in the problem
considered here, it is convenient to admit that the unit amplitude
$a=1$ should correspond to probability $P(a)=1$, from which one
obtains~$k=1$. Therefore, \eqref{eq:Born_Rule} yields the Born
Rule
\begin{equation}
P(a)=aa^{*},\label{eq:The_BR}
\end{equation}
 i.\,e. the sought-for relationship between the probability density
and the~WF amplitude.

The particle state $|\psi\rangle$ after interaction with the barrier,
according to the wave equation solution, is given by
\begin{equation}
|\psi\rangle=a|-\rangle+b|+\rangle,\label{eq:Gen_State}
\end{equation}
where $|-\rangle$ and $|+\rangle$ are the states with negative and
positive momenta, i.\,e. plane waves running to the left and to the
right from the barrier, respectively. State vectors $|-\rangle$ and
$|+\rangle$ are, actually, orthonormal eigen states of the momentum
operator. Therefore, multiplying \eqref{eq:Gen_State} by $\langle-|$
yields
\begin{equation}
\langle-|\psi\rangle=a.\label{eq:a}
\end{equation}
Now from \eqref{eq:The_BR} and \eqref{eq:a} one obtains for the
probability of realization of particle state $|-\rangle$ 
\begin{equation}
P(a)=aa^{*}=|\langle-|\psi\rangle|^{2},\label{eq:The_BR+}
\end{equation}
which is another way to put down the Born Rule. Formula \eqref{eq:The_BR+}
implies that if a particle is in state $|\psi\rangle$ then the probability
that the particle is found in state $|-\rangle$ is the squared modulus
of complex number $\langle-|\psi\rangle$, which is a projection of
the original state $|\psi\rangle$ on the final state~$|-\rangle$.
This important quantum postulate is derived here from the time-reversal
symmetry of quantum equation of motion. 

Thus, the Born Rule \eqref{eq:The_BR} or \eqref{eq:The_BR+} has
been obtained as a simple consequence of time-reversal symmetry of
quantum equation of motion. In addition to this symmetry, an important
role in the present derivation is played by the superposition principle
{[}Eq.~\eqref{eq:s}{]}, which follows from the equation of motion,
and by the assumption about the existence of indivisible particles
{[}Eq.~\eqref{eq:P(s)}{]}, which does \emph{not} follow from the
equation of motion. 

The Born Rule thus obtained for a particular case of plane wave interaction
with a potential barrier shall be valid also in the general case because
the derivation presented here is based on quantum principles ($\alpha$)
and ($\beta$) implying that the probability density $P(\psi)$ is
a \emph{universal} function applicable to any state of a particle
that can be described by a~WF.

\section{Limits of applicability of the Born Rule}

In contrast to the Schroedinger equation, the relativistic Klein-Gordon
equation contains the energy operator squared. Therefore, time reversal
does not change the Klein-Gordon equation. For this reason, under
time reversal, the Klein-Gordon equation does not require transition
to conjugate states (according to the logic set out in the Appendix).
The Born Rule, therefore, may be applicable to some particular solutions
of the Klein-Gordon equation but it should be inapplicable to the
general solution of this equation.

\section{Conclusions}

The Born Rule is derived from fundamental quantum principles, namely,
from the time-reversal symmetry and linearity of quantum equations
of motion. To obtain the Born Rule it is also necessary to admit the
existence of probability and discrete quantum particles, which is
\emph{not }a consequence of quantum equations of motion.

It should be noted that if the Klein-Gordon equation is taken as the
equation of motion in the above derivation then time-reversal~$t\rightarrow-t$
will not change the equation and, therefore, it will not require complex
conjugation of the~WF. For this reason, the Born Rule should be inapplicable
to some solutions of the Klein-Gordon equation.
\begin{acknowledgments}
The author is thankful to G.B.~Lesovik of the Landau Institute of
Theoretical Physics, Chernogolovka, and Yu.M.~Belousov of the Moscow
Institute of Physics and Technology for helpful discussions. 
\end{acknowledgments}

\section*{Appendix}

Let us consider the Schroedinger equation for an arbitrary state vector
$|\psi\rangle$: 
\begin{equation}
\hat{E}|\psi\rangle=\hat{H}|\psi\rangle,\label{eq:Sh_eq}
\end{equation}
where $\hat{E}=i\hbar\tfrac{\partial}{\partial t}$ - is the energy
operator, $\hat{H}$ - is time-independent Hamiltonian.

Upon substituting $t\rightarrow-t$ in \eqref{eq:Sh_eq} one obtains
\begin{equation}
-\hat{E}|\psi^{-}\rangle=\hat{H}|\psi^{-}\rangle,\label{eq:Not_Sh}
\end{equation}
where $|\psi^{-}\rangle=|\psi(\mathbf{r},-t)\rangle$. Equation \eqref{eq:Not_Sh}
does not coincide with the Schroedinger equation \eqref{eq:Sh_eq}
due to the \textquotedbl{}minus\textquotedbl{} sign on its left side.
In order to return to the correct equation of motion while retaining
time-reversal, it is necessary to apply the operation of Hermitian
conjugation to eq.~\eqref{eq:Not_Sh}: 
\begin{equation}
\langle\psi^{-}|\hat{E}=\langle\psi^{-}|\hat{H},\label{eq:Conj}
\end{equation}
where it is taken into account that the Hamiltonian is a Hermitian
operator $\hat{H}^{+}=\hat{H}$ while the energy operator is anti-Hermitian
$\hat{E}^{+}=-\hat{E}$ because \foreignlanguage{english}{$i\hbar\tfrac{\partial}{\partial t}$}
changes its sign under complex conjugation.

Thus, if we demand that the reversed solution should satisfy the Schroedinger
equation then time reversal of a quantum state should result in simultaneous
replacement of each ket $|\psi\rangle$ with corresponding bra~$\langle\psi^{-}|$.
This operation applied to plane wave~\eqref{eq:Plane_wave} results
in complex conjugation of amplitude~$S\rightarrow S^{*}$, which
is taken into account in eq.~\eqref{eq:Plane_wave_reversed}.

\end{document}